\documentclass{jfm}
\usepackage{graphicx}
\usepackage{epstopdf, epsfig}
\usepackage{color}
\usepackage{amssymb}
\usepackage{lipsum}
\usepackage{floatrow}
\usepackage{subfigure}

\newfloatcommand{capbtabbox}{table}[][\FBwidth]

\shorttitle{Solidification of a rivulet}
\shortauthor{A. Huerre, A. Monier, T. S\'eon, C. Josserand}

\title{
Solidification of a rivulet: \\ shape and temperature fields
}

\author{Axel Huerre\aff{1}
  \corresp{\email{axel.huerre@gmail.com}},
  Antoine Monier\aff{2},
  Thomas S\'eon\aff{2}
 \and Christophe Josserand\aff{1}}

\affiliation{\aff{1}Laboratoire d'Hydrodynamique (LadHyX), UMR 7646 CNRS-Ecole Polytechnique, IP Paris, 91128 Palaiseau CEDEX, France
\aff{2}Sorbonne Universit\'e, CNRS, UMR 7190, Institut Jean Le Rond $\partial$'Alembert,  F-75005 \ Paris, France}

\begin{document}

\maketitle

\begin{abstract}

The freezing of a water rivulet begins with a water thread flowing over a very cold surface, is naturally followed by the growth of an ice layer and ends up with a water rivulet flowing on a static thin ice wall. The structure of this final ice layer presents a surprising linear shape that thickens with the distance. 
This paper presents a theoretical model and experimental characterisation of the ice growth dynamics, the final ice shape and the temperature fields.
In a first part, we establish a 2D model, based on the advection-diffusion heat equations, that allows us to predict the shape of the ice structure and the temperature fields in both the water and the ice. 
Then, we study experimentally the formation of the ice layer 
and we show that both the transient dynamics and the final shape are well captured by the model. 
In a last part, we characterise experimentally the temperature fields in the ice and in the water, using an infrared camera. 
The model shows an excellent agreement with the experimental fields. In particular, it predicts well the linear decrease of the water surface temperature observed along the plane, confirming that the final ice shape is a consequence of the interaction between the thermal boundary layer and the free surface.

\end{abstract}



\pagebreak
\section{Problem introduction}

The processes of freezing were present at the formation of the Earth's crust \citep{Lame1831} and continues to be at the heart of the natural and industrial worlds \citep{Davis2006}. They are crucial for example in the dynamics of magmas that cool and gradually solidify until they come to rest \citep{Griffiths2000}. In this context, as often, cooling can occur from the surrounding atmosphere (or water) or from the underlying solid \citep{Huppert1989}. In ancient times, some of the lavas were even hot enough to melt the underlying rocks and shape their own thermal erosion beds \citep{Huppert1986}. 
The physical concept of freezing obviously find an infinite number of applications in the formation of ice in natural environment.
%
Among others, we can cite the sea ice formation, described as a porous matrix consisting of pure ice crystals in equilibrium with brine \citep{Wettlaufer1997}, the modelling of the dynamics of sea ice on the surface of the polar oceans \citep{Worster}, or, more commonly observed, the formation of iced rivers or lakes \citep{Beltaos2013}. The latter often involves the complex growth of stable ice covers and can be responsible for many socio-economic and ecological problems. Freezing in rivers can also give birth to waterfall ice \citep{Montagnat2010} and to surprising circle shape structures called ice circles \citep{Dorbolo2016}.


The first complete analytical treatment of the freezing of liquid is probably the Stefan's now classical work on formation of polar ice \citep{Stefan1891, Brillouin1930}, inspired by the seminal work of \cite{Lame1831}. 
In these first problems the liquid was immobile, however more recent and complex developments lead to the consideration of convective heat transfer in the ice formation context, occurring between the flowing water and the bounding ice surface \citep{Incropera2007}.
In the simplest freezing configuration of an infinite fluid layer flowing over a cold surface, significant progresses have been made in the sixties \citep{Lapadula1966, Beaubouef1967, Savino1969, Elmas1970}. 
The formation of a static frozen layer on the cold wall was predicted theoretically and the analytical solution for its thickness has been established. %
The geometry and growth of these freezing structures are governed by the balance between the heat convected by the flow, the one that diffuses in the ice and the latent heat released by the phase change. 
In these unbounded configurations however, the effect of the free surface is not taken into account. 

Indeed, the presence of a free surface can be determinant in a freezing process and solidification of capillary flows can give rise to surprising effects and striking ice structures. 
When a droplet is deposited on a cold substrate for example, the frozen drop shape and thickness depend on the contact line solidification dynamics \citep{Schiaffino1997, De-Ruiter2017} and a pointy tip is observed  \citep{Schultz2001, Marin2014, Boulogne2020}. 
If a drop impacts a cold substrate, the obtained frozen shape is the result of the complex interplay between freezing dynamics and capillary hydrodynamics  \citep{Ghabache2016, Thievenaz2019, Thievenaz2020}. The situation where ice  accretes due to multiple drop impacts can be encountered in as many practical context as ice formation on planes \citep{Cebeci2003,Baumert2018}, on bridge cables \citep{Liu2019} or on wind turbines \citep{Wang2017}. 
Finally, the solidification of a film of water flowing on a cold surface or in a cold environment \citep{Moore2017}, can also give rise to special ice structures, such as icicles \citep{Neufeld2010, Chen2011} or unstable ice ripples that form on the water-ice interface and can be observed on icicles \citep{ogawa02} or glaciers \citep{Gilpin1980, Camporeale2012}.

\begin{figure}[h!]
  \includegraphics[width=\textwidth]{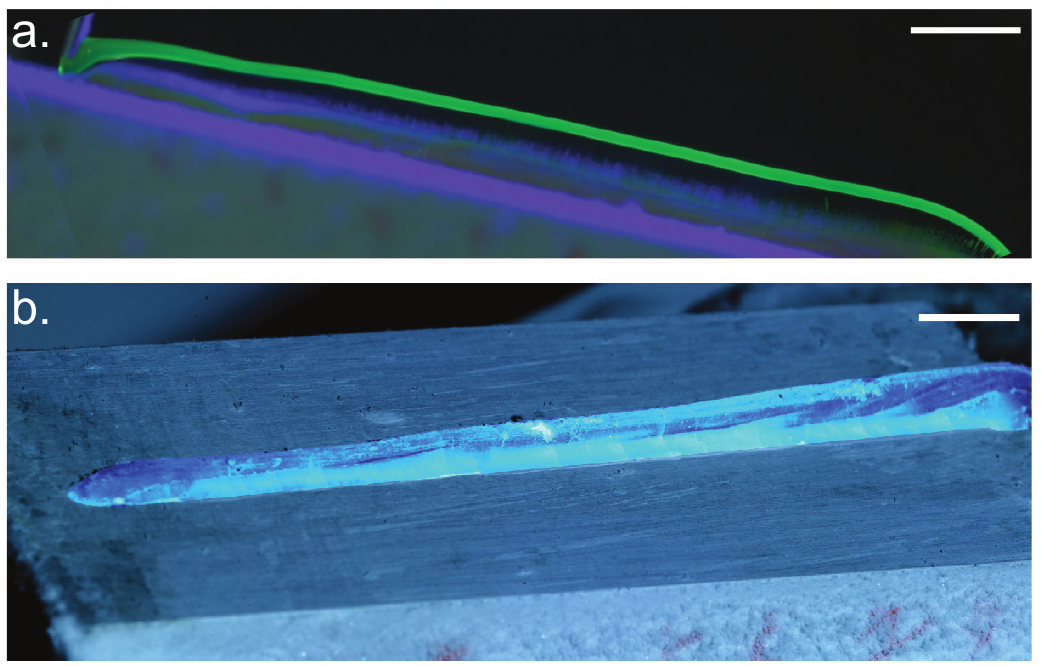}
  \caption{Frozen rivulet. (a) Experimental picture of the freezing rivulet in the steady state. The water is dyed with fluorescein and appears green under UV light. The flow goes from left to right. (b) Picture of the remaining frozen structure after we stopped the flow and cleaned the remaining water. Scale bars: 1 cm.}
\label{fig:frozenrivulet}
\end{figure}

Our study takes place in this context and focus on the solidification of a thread of water, the so-called rivulet. 
In a recent experimental study \citep{Monier2019}, we have shown that when a water rivulet flows on a cold substrate, an ice wall is growing and eventually a steady regime is reached where a thin thread of water flows on an ice structure whose thickness is almost linearly growing downstream (see Figure~\ref{fig:frozenrivulet}). In the present paper we realise a detailed study of this problem, with the derivation and solving of a complete theoretical model and the comparison with a whole range of experimental results. Consequently after the description of the experimental setup and methods that we use, the theoretical analysis is described. Then, the last section provide a complete physical analysis of the freezing rivulet. In a first part, the dynamics of the formation of the iced rivulet eventually leading to the steady regime is studied. Then, in a second part,  we investigate the temperature fields in the section of the freezing rivulet and at its free surface. In both parts, the experimental results on the shape and temperature fields are compared with the solution of the model.


\section{Experimental setup}

\begin{figure}[h!]
  \includegraphics[width=\textwidth]{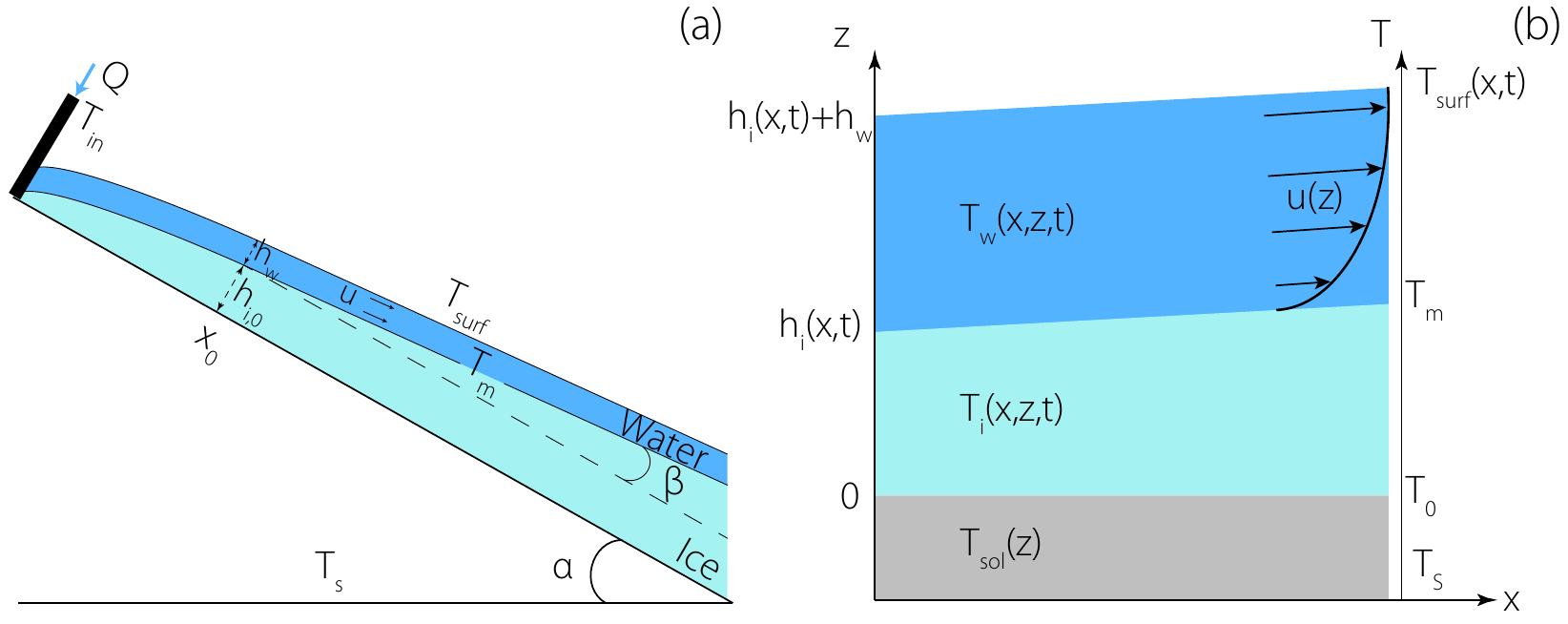}
  \caption{Notations used in the problem. (a) Lateral view of the water thread (dark blue) flowing on the ice structure (light blue). (b) Close-up schematics with the definition of the three domains on the $z$-axis, of the temperature fields and of the velocity fields.}
\label{fig:schematics}
\end{figure}

The experiment consists in flowing distilled water dyed with fluorescein at 0.5$\,$g.L$^{-1}$ along a cold aluminium block of 10$\,$cm long, with an inclination of $\alpha=30^{\circ}$ or $\alpha=60^{\circ}$ to the horizontal. The temperature of the injected water $T_{\rm in}$ ranges from 5 to $49^{\circ}$C, see Figure~\ref{fig:frozenrivulet}(a) and Figure~\ref{fig:schematics}. The water is injected through a needle (inner diameter 1.6$\,$mm) at a flow-rate $Q=20\,$mL.min$^{-1}$, such that there is no meander at room temperature \citep{LeGrand-Piteira2006}. A straight water rivulet is then formed \citep{Towell1966}, with a typical width of $2\,$mm, a thickness of $h_{\rm w}=800\,\mu$m, and a characteristic velocity of the buoyant flow $U_0\approx 10\,$cm.s$^{-1}$. 

The temperature of the aluminium substrate $T_{\rm s}$ is set by plunging the block in liquid nitrogen for a given amount of time so that it ranges from $-9$ to $-44^{\circ}$C. $T_{\rm s}$ is measured during the experiment and remains constant ($\pm 1^{\circ}$C). Experiments performed with substrate temperatures below $-44^{\circ}$C consistently lead to the fracture \citep{Ghabache2016} or the self-peeling \citep{Ruiter2018} of the ice and are not considered here.
Upon contact with the cold substrate, the water freezes and an ice layer grows while the water continues to flow on top (see Sup. Mat. movie 1).
During that process, the fluorescein concentrates between the ice dendrites, causing self-quenching and fluorescence dimming in the ice \citep{Marcellini2016}. This allows us to clearly distinguish between the ice and the water layers under UV light. The setup is placed in a humidity controlled box to avoid frost formation ($H_{\rm r} \approx 5-10\%$). 

The ice layer thickness $h_{\rm i}(x,t)$ is then measured using a Nikon D800 camera recording from the side at 30 fps. Temperature fields of the water and the ice are measured using a Flir A655sc infrared camera and by setting the average emissivity of ice and water to $\overline{\epsilon} = 0.965$. Movies of the surface temperatures (see Sup. Mat. movie 2) of the flowing water were recorded using a $25^{\circ}$ FOV, 25 mm lens ($1$ pixel $= 150\,\mu m$) while a close-up lens is added when measuring the transverse temperatures ($1$ pixel $= 50\,\mu m$, see Sup. Mat. movie 3). Image processing for both Nikon D800 camera and infrared camera is based on thresholding methods. For the ice thickness, the threshold is realised from the grey shades values. For the temperature maps, the ice-water interface is detected as follow: when temperature is below 0$^{\circ}$C, we measure temperatures in the ice, when it's above 0$^{\circ}$C, we measure temperatures in the water. Discontinuities in the temperature fields slopes set the positions of the metal-ice and the water-air interfaces.

\section{Theoretical analysis}

\subsection{General assumptions}
As the time for solidification (order of minutes) is much larger than the one of the flow (few seconds for the water to flow down the rivulet), we consider the flow to be in a (quasi)-steady regime. Consequently, both the temperature fields and the flow depend on time only through the variation of the ice layer thickness.
Moreover, the slope of the ice structure remaining small (few degrees), the liquid flow in the quasi-static regime is described considering the lubrication approximation. 
The small value of the Reynolds number ($Re=U_0h_{\rm w}/\nu \sim 80$) also ensures (with the small ice slope) that the flow is laminar and quasi-parallel. 
We may expect these approximations to fail both at very short times (where the ice layer grows rapidly) and close to the input where the ice slope can be significantly high, and 
the solution of the problem proposed below should thus be taken with care in these (small times and distances from the needle) regions. More precisely, we can estimate the time scale for the velocity profile to establish as $h_w^2/\nu \sim 0.5\,$s much smaller than the typical solidification dynamics, and the corresponding length $U_0 h_w^2/\nu \sim 5\,$cm, to a fraction of the total rivulet (about half of it).

Firstly, the water flow is considered quasi-parallel along the local ice slope (that is assuming $|h'_{\rm i}(x,t)| \ll 1$, see Figure~\ref{fig:schematics}) and the velocity field is noted $u$. The quasi-static approximation consists in neglecting the explicit time variation of the velocity and consequently the mass conservation implies that the liquid layer thickness $h_{\rm w}$ is constant \citep{Towell1966}. Within this framework, the flow is laminar and follows a semi-Poiseuille velocity profile, that can be written in the form:

\begin{equation}
u(x,z)= U_0\, \frac{z-h_{\rm i}(x)}{h_{\rm w} }\left( 2 - \frac{z-h_{\rm i}(x)}{h_{\rm w}} \right),
\label{eq:velocity}
\end{equation}
where $U_0$ is the free surface velocity. For 2D geometry, $U_0$ is obtained by the balance between the gravity and the viscous forces, yielding 
$U_0=\frac{g h_{\rm w}^2}{2 \nu} \left[\sin(\alpha)-h_{\rm i}'(x)\cos(\alpha)\right]$, although in our 3D geometry, it is a function of the lateral position in the rivulet \citep{LeGrand-Piteira2006}.
In the experiments, we used two inclination angles $\alpha$, leading to a slight increaseof $h_{\rm w}$ \citep{Towell1966} and thus $U_0$ with $\alpha$. 
Finally, let us note that even if the velocity field is along the local slope of the ice layer, we can in first approximation consider that this velocity field holds also along the $x$ direction of the substrate (the error made is of the order of the local slope  that is few percent).

\subsection{Model equations}

For the temperature field, we use the static heat equation, incorporating an advection term for the liquid domain. We further assume that the temperature of the ice-substrate interface is a time-invariant natural temperature $T_{\rm 0}$ \citep{Ruiter2018}, different from the substrate temperature $T_{\rm s}$. 
This temperature $T_{\rm 0}$ is deduced from the complete model considering the heat propagation in both the ice and the substrate \citep{Thievenaz2019}, and is a function of both the melting temperature $T_{\rm m}$ ($T_{\rm m}=0^\circ$C) and the substrate temperature far from the interface $T_{\rm s}$. 
In our case, for the substrate and temperatures considered here, one can fortuitously
approximate the interface temperature by the empirical relation:
\begin{equation}
   T_{\rm 0} \simeq 0.8\,T_{\rm s} + 0.2\,T_m 
   \label{eq:T0}
\end{equation}

The model is written for a 2D geometry ($x$ and $z$ in Figure~\ref{fig:schematics}) although the thickness of the rivulet is about half of its width. However, we expect the results of the model to give pertinent predictions for the rivulet, in particular close to its centre-line.

\begin{figure}[h!]
    \centering
    \includegraphics[width=\textwidth]{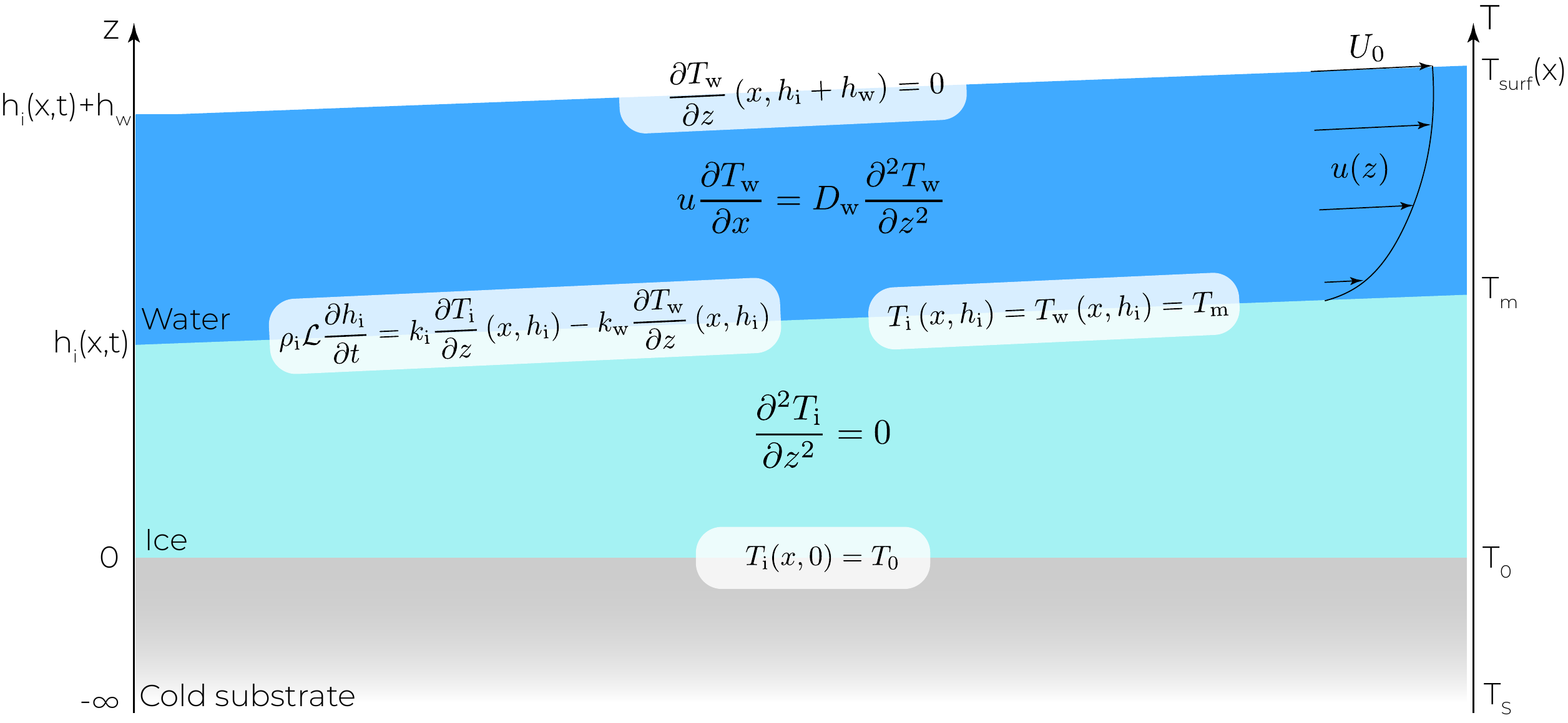}
    \caption{ Summary of the model hypotheses: a layer of ice lies between the water ($h_{\rm i}+h_{\rm w}>z > h_{\rm i}$) and the semi-infinite substrate ($z < 0$). 
    The temperature of the substrate-ice interface is set constant at $T = T_{\rm 0}$ and the one at the ice-water interface is set constant at the melting point ($T = T_{\rm m}$). The temperature in the ice and the water is given by a set of two heat equations, coupled at $z = h_{\rm i}$ by the temperature continuity and by the Stefan condition (imposing the difference of thermal fluxes to be equal to the latent heat liberated by the freezing). The velocity field in the water $u(z)$ is taken as a semi-Poiseuille with $U_0$ the velocity at the free surface $z=h_{\rm i}+h_{\rm w}$.}
    \label{fig:schem}
\end{figure}

We use thus the following set of equations for the temperature fields (see a schematic presentation of the model on Figure~\ref{fig:schem}):
\begin{itemize}
    \item in the ice, $0\leq z \leq h_{\rm i}(x,t)$, the temperature field $T_{\rm i}(x,z)$ follows
    \begin{equation}\label{eq:EDTi0}
 \frac{\partial^2 T_{\rm i}}{\partial z^2} + \frac{\partial^2 T_{\rm i}}{\partial x^2}=0
\end{equation}

\item in the water, $h_{\rm i}(x,t)\leq z \leq h_{\rm i}(x,t)+h_{\rm w}$, the temperature field $T_{\rm w}(x,z)$ follows the quasi-static advection-diffusion:

\begin{equation}
u \frac{\partial T_{\rm w}}{\partial x} = D_{\rm w} \left( \frac{\partial^2 T_{\rm w}}{\partial z^2} + \frac{\partial^2 T_{\rm w}}{\partial x^2} \right)
\end{equation}
\end{itemize}

In the latter equation, the advection term $u \partial_x T_{\rm w}$ has been taken in the $x$-direction only, consistently with the quasi-parallel flow approximation and the small slope.
Taking $Z=h_{\rm w}$ as the typical vertical lengthscale for the temperature variation, the horizontal lengthscale $X$ scales like:
$$ \frac{U_0}{X} \sim \frac{D_{\rm w}}{h_{\rm w}^2}, \;\textrm{leading to}\; X \sim Pe \,h_{\rm w},$$
where the Peclet number is defined by $Pe=U_0\,h_{\rm w}/D_w$.
From the experiments, we have the following values: $D_{\rm w} = 1.4 \cdot 10^{-7}$\,m$^2\,$s$^{-1}$; $h_{\rm w} \sim 800$\,$\mu$m; $U_0 \sim 0.1$\,m$.$s$^{-1}$. This leads to $Pe \sim 600$, demonstrating that the horizontal lenghtscale is much larger than the vertical one and that the horizontal derivatives can be neglected in the diffusion equations compared to the vertical ones.\\
The following boundary conditions have to be imposed to the temperature fields: firstly the substrate temperature $T_0$ at the interface with the ice $z=0$. Then we impose the melting temperature at the ice-water interface $z=h_{\rm i}\left(x,t\right)$. This condition corresponds to the continuity of the temperature at the ice-water interface, $T_{\rm m}$ that we take constant, neglecting its dependence with the interface curvature or velocity (Gibbs-Thomson and kinematic corrections respectively \citep{Langer1980,Worster,De-Ruiter2017,Herbaut2019}).
A zero thermal flux condition at the free surface $z=h_{\rm i}\left(x,t\right)+h_{\rm w}$, assuming that the air is insulating, since its thermal conductivity is much smaller that those of water and ice. In addition, the entry temperature $T_{\rm in}$ is prescribed at $x=0$ in the water domain.\\
Therefore, the final system of equations to be solved for determining the temperature fields both in the ice and the water domains reads:

\begin{itemize}
    \item In the ice, $0\leq z \leq h_{\rm i}(x,t)$:
\begin{equation}\label{eq:EDTi}
 \frac{\partial^2 T_{\rm i}}{\partial z^2} =0
\end{equation}

    \item In the water, $h_{\rm i}(x,t)\leq z \leq h_{\rm i}(x,t)+h_{\rm w}$:
\begin{equation}\label{eq:EDTi0f}
u \frac{\partial T_{\rm w}}{\partial x} = D_{\rm w}  \frac{\partial^2 T_{\rm w}}{\partial z^2}
\end{equation}

    \item Boundary conditions at the interfaces:
\begin{equation}
T_{\rm i}(x,0)=T_0, \hskip 0.5cm 
T_{\rm i}\left(x,h_{\rm i}\right)=T_{\rm w}\left(x,h_{\rm i}\right)=T_{\rm m},\hskip 0.25cm {\rm and} \hskip 0.25cm \frac{\partial T_{\rm w}}{\partial z} (x,h_{\rm i}+h_{\rm w})=0
\end{equation}

    \item Injection temperature in the water, $h_{\rm i}(0,t)\leq z \leq h_{\rm i}(0,t)+h_{\rm w}$:
\begin{equation}
T_{\rm w}(0,z)=T_{\rm in}
\end{equation}

\item Finally, the evolution of the ice layer thickness is described by the Stefan condition, coupling the thermal fluxes at the ice-water interface:
\begin{equation}\label{eq:Stefan}
\rho_{\rm i} \mathcal{L} \frac{\partial h_{\rm i}}{\partial t}= k_{\rm i} \frac{\partial T_{\rm i}}{\partial z} (x,h_{\rm i})-k_{\rm w} \frac{\partial T_{\rm w}}{\partial z} (x,h_{\rm i}),
\end{equation}
where $\mathcal{L}$ is the latent heat of solidification and $k_{\rm i,w}$ are the thermal conductivities of the ice and water, respectively, related to their diffusion coefficients through the general relation $k_{(\rm i,w)}=\rho_{\rm (i,w)}c_{\rm p,(i,w)}D_{\rm (i,w)}$. 
This equation is the only one containing a time derivative in our approach, a consequence of the larger time scale of ice formation compared to all the other ones. Therefore, all the temperature fields will depend on time through the (slow) time variation of $h_{\rm i}$ only. In fact, the temperature field in the solid is not {\it a priori} connected with the temperature field in the liquid film. The coupling between these two domains (liquid film and ice layer) is reduced to the Stefan condition that controls the evolution of $h_{\rm i}(x,t)$. It directly follows from the fact that the boundary condition at the ice-water interface $T=T_{\rm m}$ does not depends on time neither on $h_{\rm i}$ or other parameters in our system. Consequently, the two thermal problems (water and ice domains) are totally independent and can be solved separately, the time evolution being subjected to the variation of $h_{\rm i}$ through the Stefan equation. 
\end{itemize}

Remarkably, in the fluid domain, the set of equation and boundary conditions is similar to the so-called Graetz problem for laminar flows, written here in its 2D version \citep{Graetz1885,Fehrenbach2012}. 
Indeed, the free surface no-flux boundary condition for the temperature, equation (\ref{eq:BCair}), plays the role of a symmetry condition for a liquid flowing between two plates located at $z=0$ and $z=2\,h_{\rm w}$ while the semi-Poiseuille velocity profile follows also this symmetry.

\subsection{Dimensionless problem}
The system of equation can be made dimensionless using the two typical lengthscales introduced above $h_{\rm w} Pe$ and $h_{\rm w}$ for the horizontal and vertical directions respectively. Dimensionless temperature fields are introduced for each domain. In the water we define:

\begin{equation}
\theta_w=\frac{T_{\rm w}-T_{\rm m}}{T_{\rm in}-T_{\rm m}}\; ; \; \bar{z} = \frac{z-h_{\rm i}(x)}{h_{\rm w}} \; ; \; \bar{x} = \frac{x}{h_{\rm w}Pe}
\end{equation}
while in the ice we use: 
\begin{equation}
\theta_i=\frac{T_{\rm i}-T_0}{T_{\rm m}-T_0}\; ; \; \bar{z}_i=\frac{z}{h_{\rm w}}\; ; \;  \bar{x}=\frac{x}{h_{\rm w}Pe}\; ; \; \bar{h}_{i}=\frac{h_{\rm i}}{h_{\rm w}}
\end{equation}
The time is rescaled using the ice thermal diffusion coefficient yielding
\begin{equation}
\bar{t}=D_{\rm i} t/h_{\rm w}^2.
\end{equation}

\noindent The system of equations to be solved read then:

\begin{equation}
\frac{\partial^2 \theta_i}{\partial \bar{z_i}^2} = 0
\label{eq:adim_ice}
\end{equation}

\begin{equation}\label{eq:adim_water}
\bar{z}(2-\bar{z}) \frac{\partial \theta_w}{\partial \bar{x}} = \frac{\partial^2 \theta_w}{\partial \bar{z}^2}
\end{equation}

and

\begin{equation}\label{eq:flux}
\frac{ \partial \bar{h}_{i}}{\partial \bar{t}}(\bar{x},\bar{t})= St \left(\frac{ \partial \theta_i}{\partial \bar{z_i}}(\bar{x},\bar{h}_{i})-\frac{k_{\rm w}}{k_{\rm i}} \frac{1}{\bar{T}} \frac{\partial \theta_w}{\partial \bar{z}}(\bar{x},0) \right)\  \; , \;
\end{equation}
where the Stefan number:
\begin{equation}
    St= \frac{C_{\rm p,i} (T_{\rm m}-T_0)}{\mathcal{L}}
    \label{eq:Stefannum}
\end{equation}
compares the heat needed to vary the ice temperature from the substrate to the melting one with the latent heat of solidification.

\noindent The boundary conditions read:
\begin{equation}\label{eq:bcT0}
\theta_w(0,\bar{z} ) =1 \; ; \; \textrm{$T_{\rm in}$ at the entrance}
\end{equation}

\begin{equation}\label{eq:BCTm}
\theta_w(\bar{x},0) = 0  \; ; \; \textrm{$T_{\rm m}$ at the water-ice interface}
\end{equation}

\begin{equation}\label{eq:BCair}
\frac{ \partial \theta_w}{\partial \bar{z}}(\bar{x},1)= 0  \; ; \; \textrm{Insulating water-air interface - no flux}
\end{equation}

\begin{equation}\label{eq:BCTiTs}
\theta_i(\bar{x},0) =0 \; ; \; \textrm{$T_0$ at the metal-ice interface}
\end{equation}

\begin{equation}\label{eq:BCTiTm}
\theta_i(\bar{x},\bar{h}_{i}) = 1  \; ; \; \textrm{$T_{\rm m}$ at the water-ice interface}
\end{equation}

\noindent Finally, the reduced temperature 
\begin{equation}
    \bar{T}=\frac{T_{\rm m}-T_0}{T_{\rm in}-T_{\rm m}}
    \label{eq:tbar}
\end{equation}
plays the role of the control parameter of the problem.

\subsection{Model solution}
\label{ModelSolution}

As explained, the two thermal equations (\ref{eq:adim_water}) and (\ref{eq:adim_ice}) can be treated independently, since the variations of $\bar{h}_{\rm i}(\bar{x},t)$ do not intervene explicitly. 

\noindent Firstly, the resolution of the equation is straightforward in the ice layer, yielding:

\begin{equation}
\theta_i= \frac{\bar{z}_i}{\bar{h}_{i}}=\frac{z}{h_{\rm i}(x,t)}.
\label{eq:thetai}
\end{equation}

\noindent On the other hand, the temperature field in the water can be deduced from a 2D Graetz problem and we recall here the main properties of this solution \citep{Incropera2007}. 
Using the linearity of equation (\ref{eq:adim_water}) and separation of variables, we can exhibit a quasi-analytical solution of the problem.

More precisely, we seek solutions of the equation for $0\le \bar{z} \le 1$ in the form:

\begin{equation}
\theta_w(\bar{x},\bar{z})= \sum_{n=1}^{\infty} \theta_n(\bar{x},\bar{z})=\sum_{n=1}^{\infty} A_n \Phi_n(\bar{z}) e^{-\lambda_n^2 \bar{x}},
\label{eq:PhiGen}
\end{equation}
where $\Phi_n$ and $\lambda_n$ are the eigenfunctions and the eigenvalues, respectively, of the following Sturm-Liouville problem, composed of the equation:

\begin{equation}
\Phi_n''=-\lambda_n^2 \, \bar{z} \, (2-\bar{z})\, \Phi_n 
\label{eq:phi}
\end{equation}
and the boundary conditions:
\begin{equation}
 \Phi_n(0)=0\; {\rm and} \; \Phi_n'(1)=0
 \label{eq:bc}
\end{equation}

 \begin{figure}[b!]
    \includegraphics[width=0.9\textwidth]{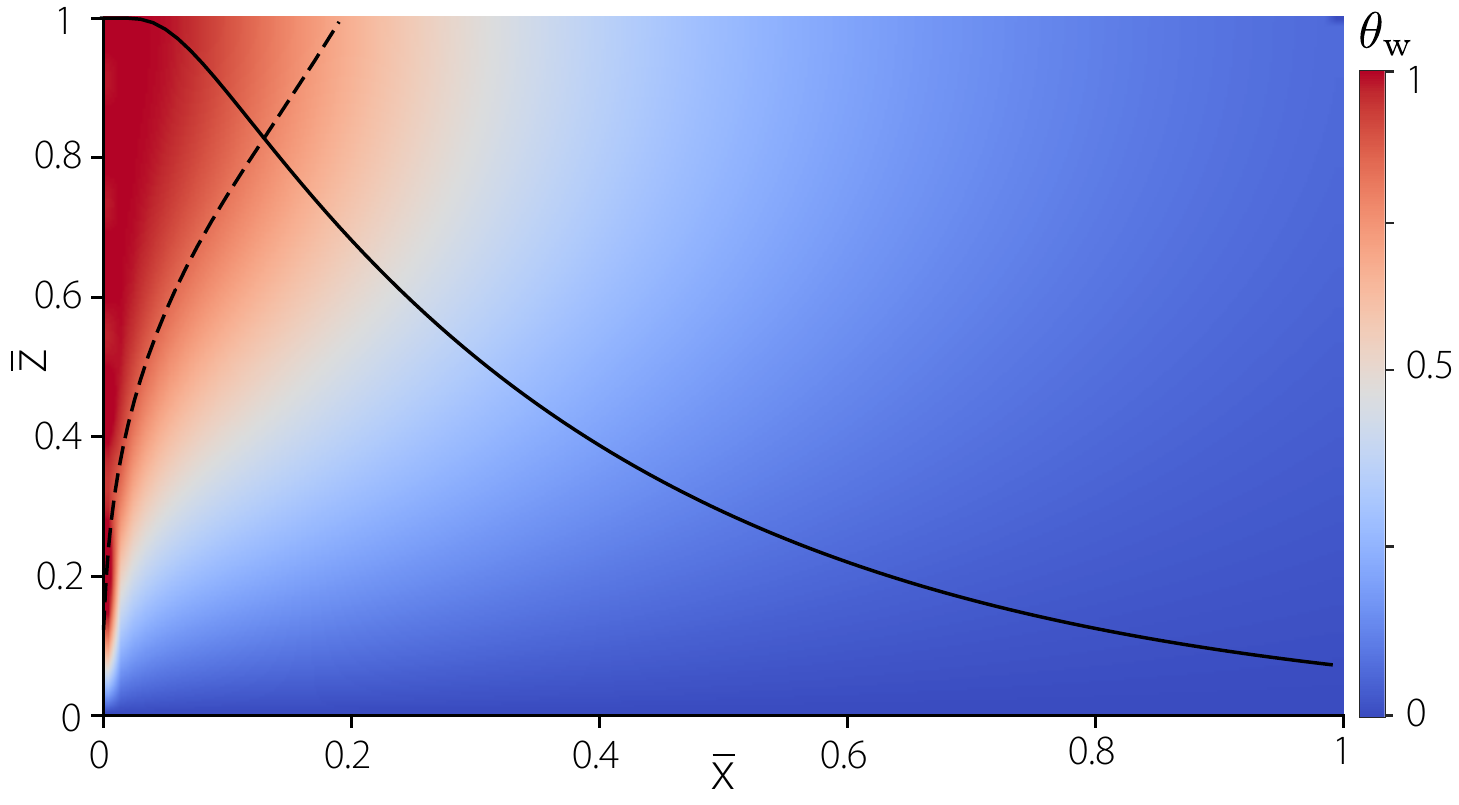}
    \caption{Colour map of the temperature field in the liquid layer, solution of equation (\ref{eq:adim_water}) with the associated boundary conditions. The solution is shown in dimensionless unit, and the colour scale is indicated on the right. The dashed line shows the growth of the thermal boundary layer defined in equation (\ref{eq:Th-BL}).
    The solid line represents the temperature at the free surface, $\theta_{\rm surf}(\bar{x})=\theta_w(\bar{x},1)$.}
    \label{fig:Tmap}
\end{figure}

The eigenfunctions $\Phi_n$ involve cylindric functions and the boundary conditions lead to a discrete infinite set of positive eigenvalues $\lambda_n$.
To evaluate the coefficients $A_n$ in equation (\ref{eq:PhiGen}), we apply the last boundary condition, equation (\ref{eq:bcT0}), which specifies the temperature at the inlet. Using the orthogonal property of the Sturm-Liouville system, we obtain:
 \begin{equation}
     A_n=\int^1_0 \Phi_n(\bar{z}) \,\bar{z} \,(2-\bar{z}) d\bar{z}
 \end{equation}
The coefficients $A_n$ and $\lambda_n$ as well as the functions $\Phi_n$ are provided in Appendix.
Figure~\ref{fig:Tmap} shows the temperature map of the solution: as $\bar{x}$ increases, we can observe the growth of the thermal boundary layer starting from $\bar{z}=0$ at $\bar{x}=0$ due to the cooling of the liquid from the plate, as the liquid flows. The size of this boundary layer 
\begin{equation}
\delta(\bar{x})=\frac{1}{\displaystyle \frac{\partial \theta_w}{\partial \bar{z}}(\bar{x},0)},
\label{eq:Th-BL}
\end{equation}
defined as the inverse of the temperature slope at $\bar{z}=0$ is indicated on the map as a black dashed line. 
At small scale, it can be shown that $\delta(\bar{x}) \propto \bar{x}^{1/3}$ and we observe that the boundary layer reaches in fact the liquid film thickness for $\bar{x} \sim 0.2$. Further on, we observe that the surface temperature 
$\theta_w(\bar{x},1)$ decreases rapidly, as illustrated by the solid line on Figure~\ref{fig:Tmap}.

Finally, one can show from the equation \ref{eq:PhiGen} that the solution of the equation is well approximated by its first mode already for $\bar{x}>0.05$, leading to:
\begin{equation}
\theta_w(\bar{x},\bar{z}) \sim A_1\, \Phi_1(\bar{z})\, e^{-\lambda_1^2 \bar{x}}
\label{ap:thetaw}
\end{equation}
and predicting an exponential decrease of the surface temperature for $\bar{x} \geq 0.05$, as observed in Figure~\ref{fig:Tmap}.
For the experiments, it means that this single mode temperature field is valid for  $x \geq 0.05 \, h_{\rm w} \, Pe \sim 2$ cm, which is a fraction of the substrate length.

\section{Physical analysis}

Using our experimental tools and the theoretical model proposed before, we can now proceed to a complete physical analysis of the freezing rivulet. We start by studying the ice layer growth.

\subsection{Formation of the ice layer}

\subsubsection{General description of the rivulet growth} 

\begin{figure}[h]
\centering
\includegraphics[width=1\textwidth]{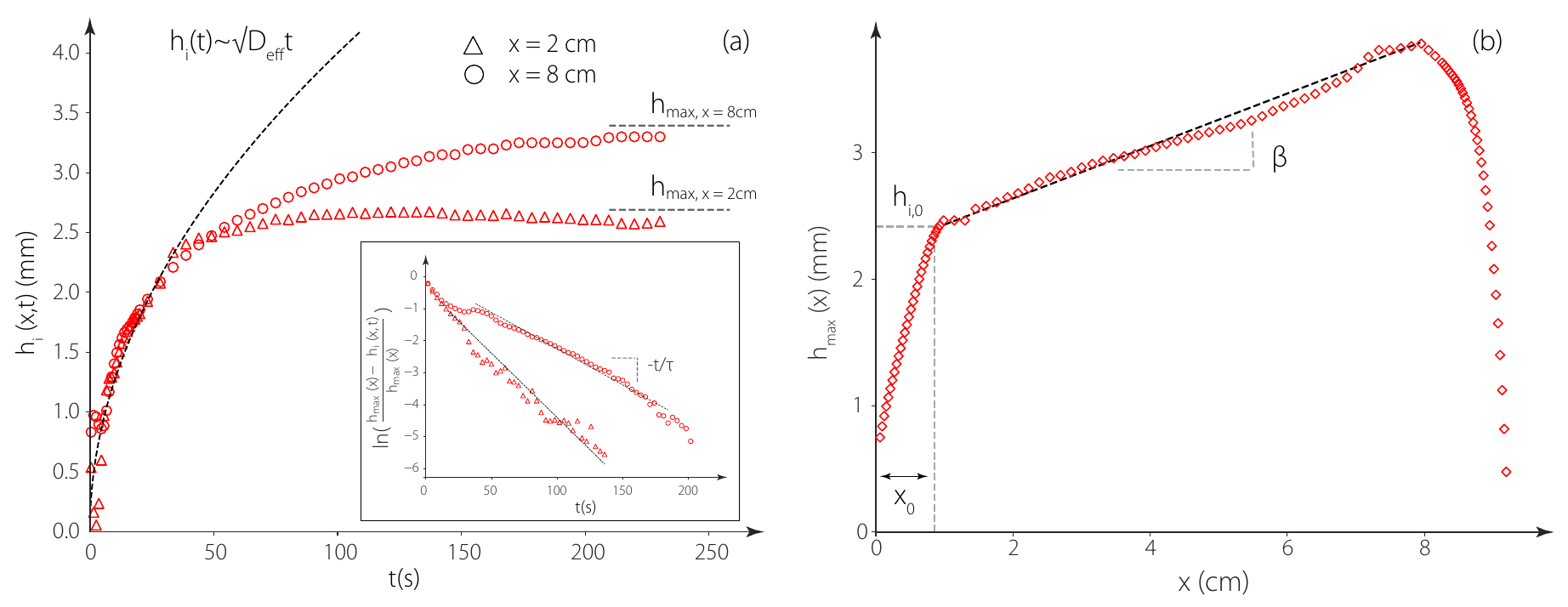}
\caption{Ice structure growth. (a) Ice thickness over time for two different plane positions ($T_{\rm in} = 11^{\circ}$C, $T_0 = -11.4^{\circ}$C). The ice growth is first diffusive and then converges exponentially towards a maximum value $h_{\rm max}(x)$. Inset: $(h_{\rm max}(x)-h_{\rm i}(x,t))/h_{\rm max}(x)$ as function of the time in a lin-log scale. $\tau$ is defined as the characteristic time-scale. (b) Profile of the frozen rivulet at the end of the experiment, for the same parameters. After an \textit{entry} zone of length $x_{\rm 0}$, the ice thickness profile is linear, characterised by its angle $\beta$.}
\label{FigPRL}
\end{figure}

Figure~\ref{FigPRL}(a) presents the ice layer thickness as a function of time for two different positions along the substrate ($x=2$ and $8\,$cm), for $T_{\rm in} = 11^{\circ}$C and $T_0 = -11.4^{\circ}$C ($\bar{T}=-T_0/T_{\rm in}=1.04$). As shown in a previous study \citep{Monier2019}, at early times the ice layer grows homogeneously along the plane following a diffusive dynamics $h_{\rm i}(t)=\sqrt{D_{\rm eff}t}$. The height profiles are parallel to the substrate. $D_{\rm eff}$ is an effective diffusive coefficient, solution of a transcendental equation that involves $T_{\rm s}$, $\mathcal{L}$ and the ice and substrate thermal coefficients \citep{Thievenaz2019}.

After this diffusive regime, we observe a second regime where the ice layer continues to grow until it reaches a maximum height $h_{\rm max}(x)$ that increases along the plane. 
The inset of Figure~\ref{FigPRL}(a) presents the same data set but using a logarithmic scale for the normalised difference to the final ice height $(h_{\rm max}(x)-h_{\rm i}(x,t))/h_{\rm max}(x)$. It shows that, in this second regime, the ice thickness converges exponentially towards its asymptotic value. We then define the characteristic growth time $\tau$ by fitting the ice height for large times, following $(h_{\rm max}(x)-h_{\rm i}(x,t))/h_{\rm max}(x) \sim A(x)e^{-t/\tau}$. 

Following this second phase, the ice layer stops growing and the system reaches a permanent regime consisting of a static ice structure, of thickness $h_{\rm max}(x)$, on top of which a water thread is flowing. 
Such a permanent regime can be understood qualitatively by considering the thermal fluxes at the ice-water interface where the temperature is always at the melting one (0$^\circ$C). 
The water is dispensed at a constant temperature, inducing the development of a thermal boundary layer that imposes a temperature gradient perpendicular to the ice-water interface, constant in time. 
On the other side of the interface, the temperature gradient in the ice can be estimated with $(T_{\rm m}-T_0)/h_{\rm max}(x)$. 
At early times, the ice layer is very thin and the gradient is high. Consequently, the ice grows in order to decrease its temperature gradient and stops when the heat fluxes on both sides of the ice-water interface balance.
Figure~\ref{FigPRL}(b) presents the maximum height reached by the ice layer, $h_{\rm max}(x)$, as a function of $x$ . 
After a first \textit{entry} zone, characterised by a steep ice thickness increase, $h_{\rm max}(x)$ is well described by a line of slope $\beta$ as illustrated by the dashed line: $h_{\rm max}(x)=h_{\rm i,0}+\beta (x-x_0)$.
In the following, we will show that this linear height profile is due to the heat balance between the ice layer and the liquid rivulet, and is deeply related to the fact that the thermal boundary layer has reached the rivulet thickness $h_{\rm w}$.

\subsubsection{Theoretical permanent ice thickness profile}

Knowing the temperature field in both domains (water and ice), we can write the evolution equation for the ice layer thickness. This equation is valid within the quasi-static approximation made here, where the time variations of the fields are simply subjected to the evolution of $\bar{h}_i(\bar{x},\bar{t})$.
In this context, the Stefan equation (\ref{eq:flux}) leads to the equation for $\bar{h}_i(\bar{x},t)$:

\begin{equation}\label{eq:film}
\frac{ \partial \bar{h}_i}{\partial \bar{t}}(\bar{x},\bar{t})= St \left(\frac{ 1 }{\bar{h}_i}-\frac{k_{\rm w}}{k_{\rm i}} \frac{1}{\bar{T} }\frac{\partial \theta_w}{\partial \bar{z}}(\bar{x},0) \right)\  \; , \;
\end{equation}
where we have used the constant gradient solution from equation (\ref{eq:thetai}) for the temperature in the ice layer. 
$\theta_w$ is then taken as the general solution for the water temperature field, that we might approximate by the first mode approximation for $\bar{x}>0.05$, given by equation (\ref{ap:thetaw}). We can thus deduce the final shape of the ice layer, valid in our experiment for $x>2.5\,$cm:

\begin{equation}
h_{\rm max}(x) = h_{\rm w} \,\bar{T} \, \frac{k_{\rm i} }{k_{\rm w}} \frac{1}{A_1}  \, \frac{1}{\Phi_1' (0)} \, \exp\left( \lambda_1 ^2 \frac{x-x_0}{h_{\rm w} Pe}\right)
\label{hmax}
\end{equation}

In the next two sections, we study how the rivulet reach this permanent regime  and how this theoretical prediction compares with the experimental measurements.

\subsubsection{Convergence to the static shape}

Following the initial diffusive growth, we wonder how the ice dynamics reaches its static shape. We develop an asymptotic analysis around the stationary state, $h_{\rm i}(x,t)=h_{\rm max}(x) \, (1+\epsilon\,f(t))$, with $\epsilon \ll 1$. Substituting in equation (\ref{eq:film}) gives a differential equation for $f$:

\begin{equation}
    f'(t)=\frac{-St}{D_{\rm i}\, h_{\rm max}^2}\,f(t)
\end{equation}

It confirms the exponential behaviour of the ice growth towards its static shape that we observed experimentally (see Figure~\ref{FigPRL}(a)). Subsequently, we can determine the characteristic time $\tau$ using the theoretical prediction of the final ice shape given by equation (\ref{hmax}):

\begin{equation}
    \tau=\frac{h_{\rm max}^2}{D_{\rm i}\,St} = \frac{h_{\rm w}^2}{D_{\rm i}}\frac{\bar{T}^2 }{St}\, \left(\frac{k_{\rm i} }{k_{\rm w}} \frac{1}{A_1}  \, \frac{1}{\Phi_1' (0)}\right)^2 \, \exp \left(2\,\lambda_1 ^2\, \frac{x-x_0}{h_{\rm w} Pe}\right)
    \label{eq:tauth}
\end{equation}

\begin{figure}[h!]
\centering
  \includegraphics[width=0.75\textwidth]{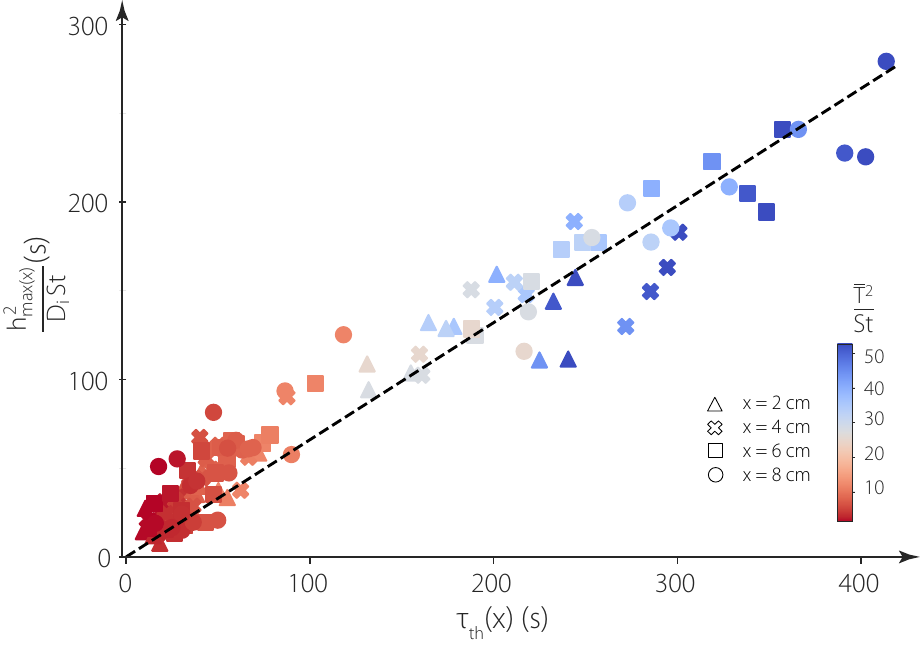}
  \caption{Experimental convergence time $h_{\rm max}^2/(D_{\rm i}\,St)$ against theoretical convergence time as predicted by equation (\ref{eq:tauth}), for a plane inclination of $\alpha = 30^{\circ}$. Markers stand for the positions along the plane and colour-code for $\bar{T}^2/St$.}
\label{tau_exp-vs-tau_thq}
\end{figure}

Figure~\ref{tau_exp-vs-tau_thq} presents the convergence time measured on the experimental curves (such as in the inset of Figure~\ref{FigPRL}) as a function of this theoretical prediction. For the different physical parameters, we use $k_{\rm i}=2100$\,W.K$^{-1}$ and $k_{\rm w}=580$\,W.K$^{-1}$, $h_{\rm w}$=800\,$\mu$m, $D_{\rm i} = 1.2\cdot 10^{-6}\,$m$^{2}.$s$^{-1} $, $c_{\rm p,i} = 2090\,$J.kg$^{-1}$.K$^{-1}$, $\mathcal{L} = 3.3 \cdot 10^{5} \,$ J.kg$^{-1}$, and $Pe=570$. $\lambda_1$, $A_1$, and $\Phi_1'(0)$ are numerically evaluated as $1.68$, $0.78$, and $2.2$ respectively.
This comparison is particularly convincing, since the coefficient between the two times is always around 0.7, very close to 1. It provides a first validation of the theoretical prediction of the static ice shape $h_{\rm max}$ and consequently of the entire model that led to this prediction. Moreover, it allows us to forecast the characteristic time it takes for the rivulet to reach its maximum height. Finally, it predicts that this time scales as $\bar{T}^2 /St \propto (T_{\rm m}-T_{\rm 0})/(T_{\rm in}-T_{\rm m})^2$ indicating that the variations of water temperature is the main parameter to control the convergence time.

Although the agreement between the model and the experiments is very good with regards to the approximations made, we observe some discrepancies (the fit of the data gives a slope $0.7$ instead of $1$ in Figure~\ref{tau_exp-vs-tau_thq}, the results are spread) that can be attributed to different factors. Firstly, we have taken a constant $h_{\rm w}$ over the experiments while it varies slightly with the temperatures, witnessing probably a variation of the wetting properties of the water on ice. Moreover, since the experiments were performed with non-degassed water, we always observe bubbles in the part of the ice close to the substrate (see for example the difference of colour in the ice on Figure \ref{fig:frozenrivulet}(a)). The effect of bubbles in the ice during experiments of freezing of capillary objects is currently investigated \citep{Chu2019} and composite models can be used to predict the conductivity when knowing the concentration, organisation and size of the bubbles in the ice \citep{Wegst2010}. Typically, a reduction of a factor two in the conductivity would correspond to 20-30\% of air in the ice, consistent with the typical front velocity at the beginning of our experiment $\sim$5\,mm/min \citep{Carte1961}. We thus believe that an effective thermal conductivity $k_{\rm i}$ (or even varying with space) should be considered in the ice to account more finely for the experimental observations.

\subsubsection{Experimental permanent ice thickness profile}

After the two regimes of growth fully characterised before: a diffusive one followed by an exponential relaxation, the ice reaches a static shape. Figure~\ref{fig:frozenrivulet} shows two examples of such shape and Figure~\ref{FigPRL} (b) is a plot the corresponding ice thickness. As we saw, the experimental profile is linear and can be written $h_{\rm max}(x) = h_{\rm i,0} + \beta\,(x-x_0)$.

\noindent Theoretically, this thickness profile is given by equation (\ref{hmax}), where the experimental values of x ($\bar{x}>0.05$, that is $x>2.5$ cm) allow us to consider only the first mode and to expand the exponential term. We thus recover the linear relationship between $h_{\rm max}$ and $x$ :


\begin{eqnarray}\label{eq:hmax_hio}
h_{\rm max}(x) & \sim & h_{\rm w} \,\bar{T} \, \frac{k_{\rm i} }{k_{\rm w}} \frac{1}{A_1}  \, \frac{1}{\Phi_1' (0)} \, \left(1+ \lambda_1 ^2 \frac{x-x_0}{h_{\rm w} Pe}\right) \\[16pt]
  & = & 1.7 \cdot 10^{-3} \,\bar{T}\,+ 1.1 \cdot 10^{-2}\,\bar{T}\,(x-x_0)
\end{eqnarray}

\begin{figure}[ht!]
\centering
  \includegraphics[width=1\textwidth]{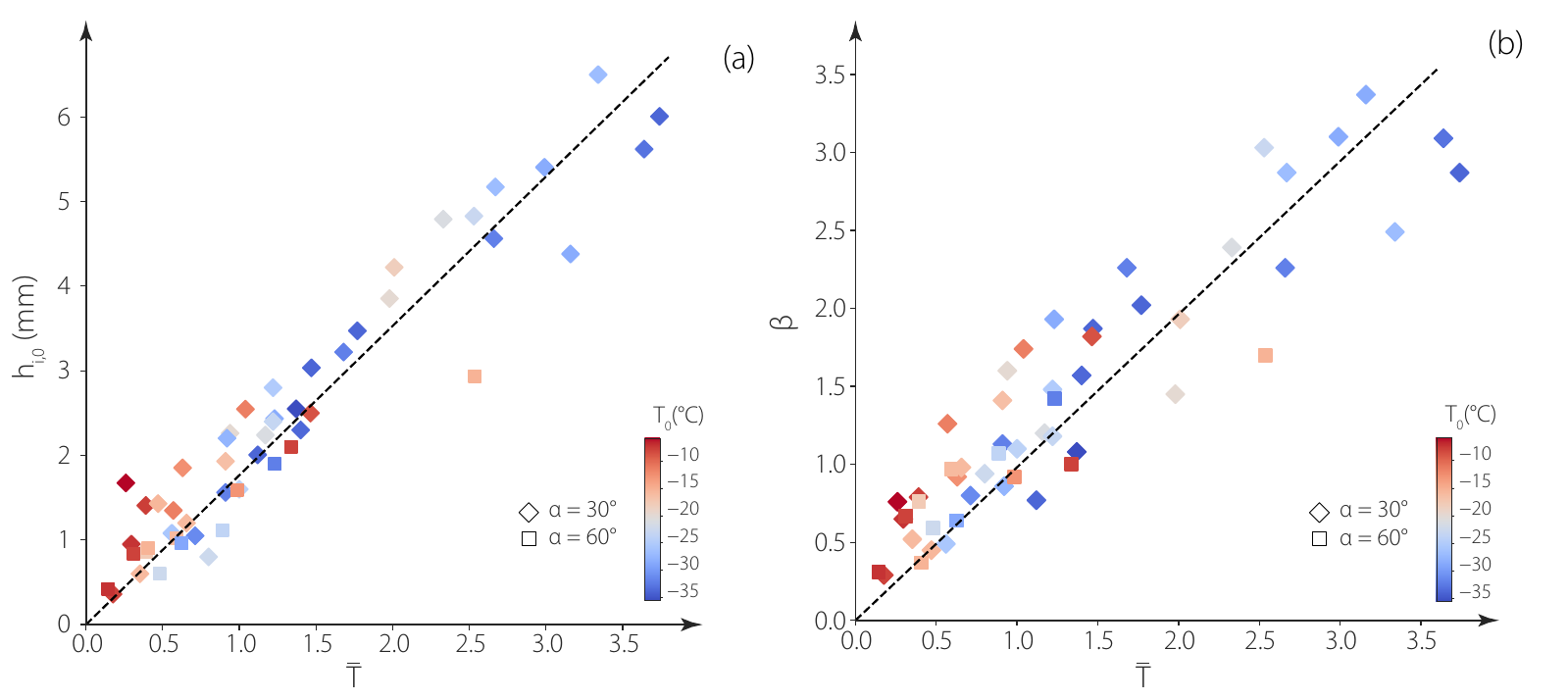}
\caption{Permanent ice structure profile. (a) Measured $h_{\rm i,0}$ against $\overline{T}$, the dashed line is the best fit of the data: $1.8\cdot 10^{-3}\, \overline{T}$. Colours stand for the substrate-ice interface temperature $T_{\rm 0}$ and markers for the plane inclination. (b) Measured slope $\beta$ of the ice profile against $\overline{T}$. Colours stand for the substrate-ice interface temperature and markers for the plane inclination.}
\label{fig:MaxThicknessProfile}
\end{figure}

Figure~\ref{fig:MaxThicknessProfile}(a) shows the thickness of the ice at the beginning of the linear profile $h_{\rm i,0}$ obtained on the experimental profiles, as a function of the reduced temperature $\overline{T}$. The data clearly exhibit a linear trend and a fit, represented as a dashed line on Figure~\ref{fig:MaxThicknessProfile}(a), gives a coefficient $1.8\cdot 10^{-3} \,$m. this value, very close to the prediction of $1.7\cdot 10^{-3}\, $m highlights the very good performance of the model. In  Figure~\ref{fig:MaxThicknessProfile}(b), the experimental values of the slope $\beta$ are plotted as a function of $\bar{T}$. Again, the linear behaviour is recovered and the dotted line around which all the data gather is $\beta=1.7\cdot 10^{-2}\,\bar{T}$, of the same order then the theoretical one. Overall, these two plots show that the model is well suited to predict the maximum ice thickness profile, even though we had to expand the exponential term. It is interesting to note that if at short distance the ice is linear, a long rivulet would then have a exponential shape and would thus grow very fast, forming a surprisingly high ice structure. \\

\subsection{Temperature fields}

\subsubsection{Transverse temperature profiles}

We were able for few experiments to measure the temperature maps of the rivulet while flowing, in a small region of the plane. This region is situated between $x= 2\,$cm and $x=5\,$cm, that is roughly at the end of the thermal boundary layer regime and the beginning of the free surface one (where the thermal boundary layer thickness is of the order of the rivulet height $h_{\rm w}$). The infrared camera resolution of 50$\,\mu$m allows us to record around 20
measurement points along $z$ in the water and between 20 and 60 in the ice.
Keeping in mind that the camera records the temperature at the lateral surface of the rivulet, we compare in the following the experimental results with the theoretical expressions of the temperatures in the ice and in the water obtained in Section \ref{ModelSolution}.

\begin{figure}[h!]
\centering
  \includegraphics[width=1\textwidth]{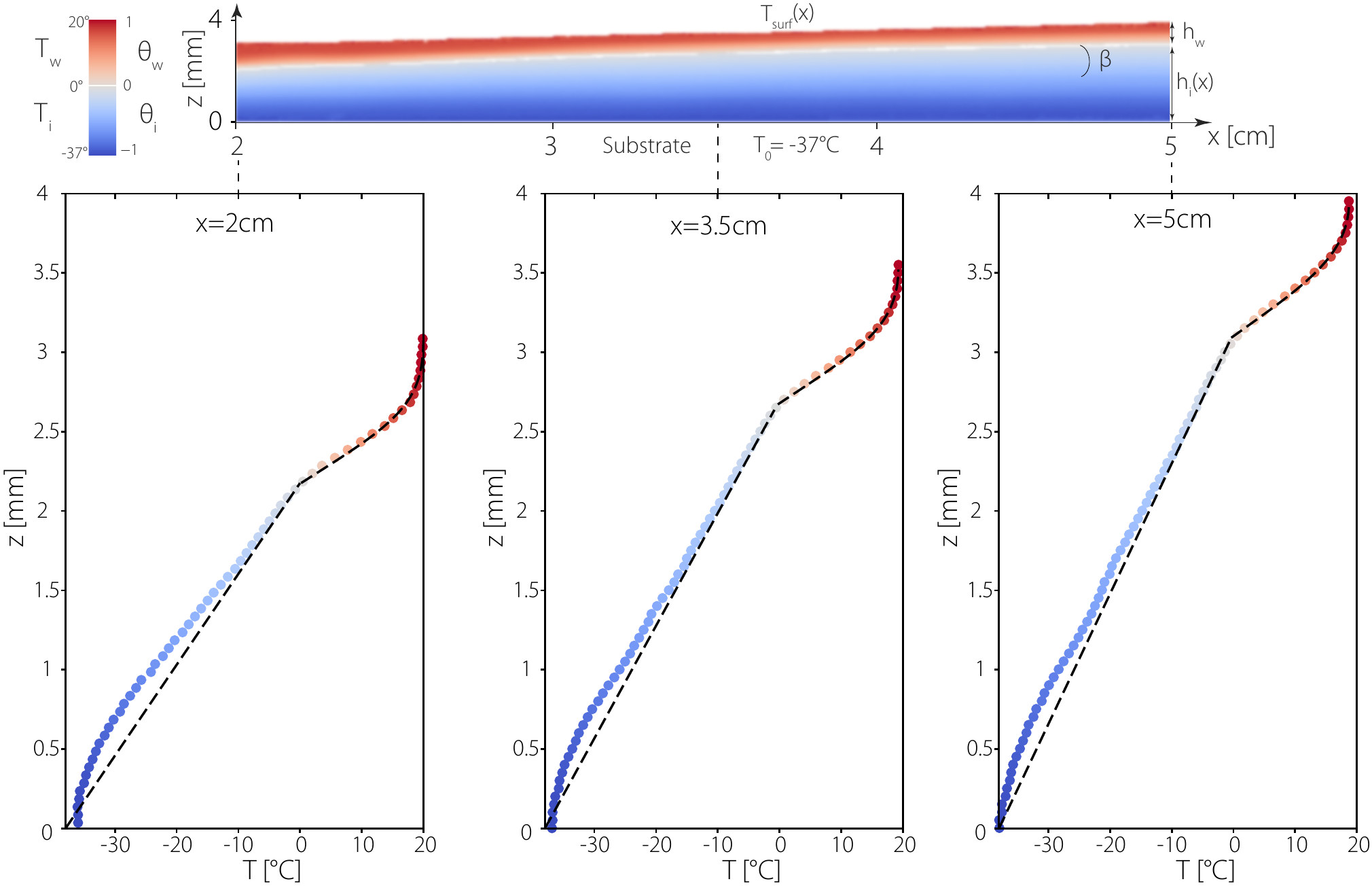}
  \caption{Cross-section temperature profiles and model. Top: Experimental picture of the temperature fields in the water (red) and in the ice (blue). $z=0$ corresponds to the substrate-ice interface. Bottom: Experimental and theoretical profiles (equations (\ref{eq:thetai}) and (\ref{eq:PhiGen})). To plot the theoretical temperature profiles, the velocity $U_0$ was adjusted to 19\,cm$\cdot$s$^{-1}$ (consistently with experimental estimates), the inlet temperature was set to $T_{\rm in}=20^\circ$C, the water thickness $h_{\rm w}=0.9\,$mm to its measured value and the origin of the x-axis was shifted of 3.8 cm.
  }
\label{fig:Tz}
\end{figure}

The temperature field shown at the top of Figure~\ref{fig:Tz} is obtained with the infrared camera placed on the side of the rivulet in the permanent regime. The temperature is colour-coded according to the colour bar in the top left corner of the figure.  
The line $T=T_{\rm m}=0^\circ$C is represented in white and corresponds to the ice-water interface. The water, appearing in red, is flowing from left to right, on the ice, appearing in blue. The ice structure has reached its static shape, it is not growing anymore. We thus recognise the angle $\beta$ formed by the ice structure. A measurement on this picture gives $\beta=1.7^\circ$, consistent with the prediction from Figure~\ref{fig:MaxThicknessProfile}(b), considering that $\bar{T}=1.85 $ in this experiment.
We observe on this map the transverse temperature gradient across the whole structure: the temperature increases from $T_{\rm 0}=-37^\circ$C at the ice-substrate interface to $T=T_{\rm m}=0^\circ$C at the ice-water interface, and up to the water-air surface temperature, that is close to  $24^\circ$C, the injection temperature of the water $T_{\rm in}$. The water-air surface temperature will be discussed later.

To go further in the quantitative analysis, temperature profiles can be extracted from this temperature field. The three graphs on Figure~\ref{fig:Tz} show the temperature profiles, corresponding to the temperature map above, at three different positions on the plane: $x=$2, 3.5 and 5\,cm.  The experimental points are colour-coded in the same way as in the map. The origin of the $z$-axis is taken on the metallic plane. As a consequence, the ice-water interface, localised by the discontinuous temperature gradient, is at a different height in each graph.

The experimental temperature profile in the ice is mainly linear except in a thin zone close to the substrate ($z\leqslant0.5\,$mm). This deviation from the linear profile is a signature of the temperature field in the substrate \citep{Thievenaz2019}.
Consequently, the linear model given by equation (\ref{eq:thetai}) and plotted with a dashed line is very satisfying close to the ice-water interface and becomes more approximate near the substrate. Note that the substrate temperature measured with a thermocouple directly on the metal is $T_{\rm s}=-44^\circ$C and the temperature of the ice in contact with the substrate is $T_{\rm 0}=-37^\circ$C. This difference is perfectly consistent with the empirical relation between $T_{\rm 0}$ and $T_{\rm s}$ given by equation ($\ref{eq:T0}$) that we deduced from a complete model considering the heat propagation in both the ice and the substrate.

In the water, the experimental temperature profiles, appearing in red, highlights the thermal boundary layer behaviour: the temperature is almost constant in a zone close to the free surface. We also notice on the profiles that this zone reduces in size as we go downstream.
The theoretical temperature profiles given by equation (\ref{eq:PhiGen}) are plotted with dashed lines on the same graphs. They superimpose perfectly to the experimental profiles in the three cases. Therefore, the model precisely reproduces the flux at the ice-water interface, the surface temperature and the variation in the water bulk.

Finally, we clearly observe a discontinuity in the temperature slopes at the ice-water interface, testifying the difference of conductivities of the two media (equation (\ref{eq:Stefan})). 
A measurement of the experimental temperature slopes, with a linear fit on the 9 first points in each phase, gives $k_{\rm i}/k_{\rm w}=2.34$. 
This ratio should be equal to 3.6 by taking the conductivity values for pure ice and water. As we discussed before, we attribute the difference between our experimental measurement of $k_{\rm i}/k_{\rm w}$ and the one given by the pure body values to the variation of thermal conductivity in ice due to the inclusion of bubbles. Consequently, by taking the conductivity of pure water, this experimental measurement can give us an estimation of the thermal conductivity of our ice in presence of bubbles, we find $k_{\rm i}=1360$\,K/m. Interestingly, this effect could explain the discrepancy observed in Figure~\ref{tau_exp-vs-tau_thq} for the experiments that last long (blue points). For these long experiments, we consistently observed a very thick ice layer with a big zone full of bubbles and one free of bubbles. We can thus expect the thermal conductivity in these experiments to be smaller and the theoretical convergence time to be reduced consistently.

\subsubsection{Comparison of the convective to conductive heat transfer}

This section aims at comparing the heat transfer by convection from the ice-water interface to the moving liquid and by pure conduction (diffusion) to a hypothetically motionless liquid. This demands to compute  the Nusselt number of our experiment. 
It is here defined as the ratio of the conductive to the convective thermal resistance of the fluid~\citep{Incropera2007} : 

\begin{equation}
    Nu= 4\;\frac{\displaystyle\frac{\partial T_{\rm w}}{\partial z}(z=h_{\rm i})}{\displaystyle\frac{<T>-T_{\rm m}}{h_{\rm w}}}
\end{equation}

\noindent 
We define the mean temperature $<T>$ by expressing the rate at which thermal energy is advected with the fluid ($\rho <u> h_{\rm w}c_{\rm p} <T>$) with integration over the cross section:

\begin{equation}
   <T> =\frac{\displaystyle\int_{0}^{h_{\rm w}}u\, T\, {\rm d}z}{h_{\rm w}\, <u>}
   \label{eq:Tmoy}
\end{equation}

\noindent where $ <u> = 2U_0/3$ is the flow mean velocity. This amounts to consider the average temperature weighted by the fluid velocity.
To do so, the numerical integration of the advected experimental temperature profile is performed assuming the theoretical half-Poiseuille velocity field used in the previous analysis.
Moreover, as discussed in the previous section, the conductive thermal flux at the ice-water interface ($\partial T_{\rm w}/\partial z (z=h_{\rm i})$), can be deduced from the experimental temperature profiles, for different positions along the plane. 

\begin{figure}[h!]
\centering
  \includegraphics[width=0.6\textwidth]{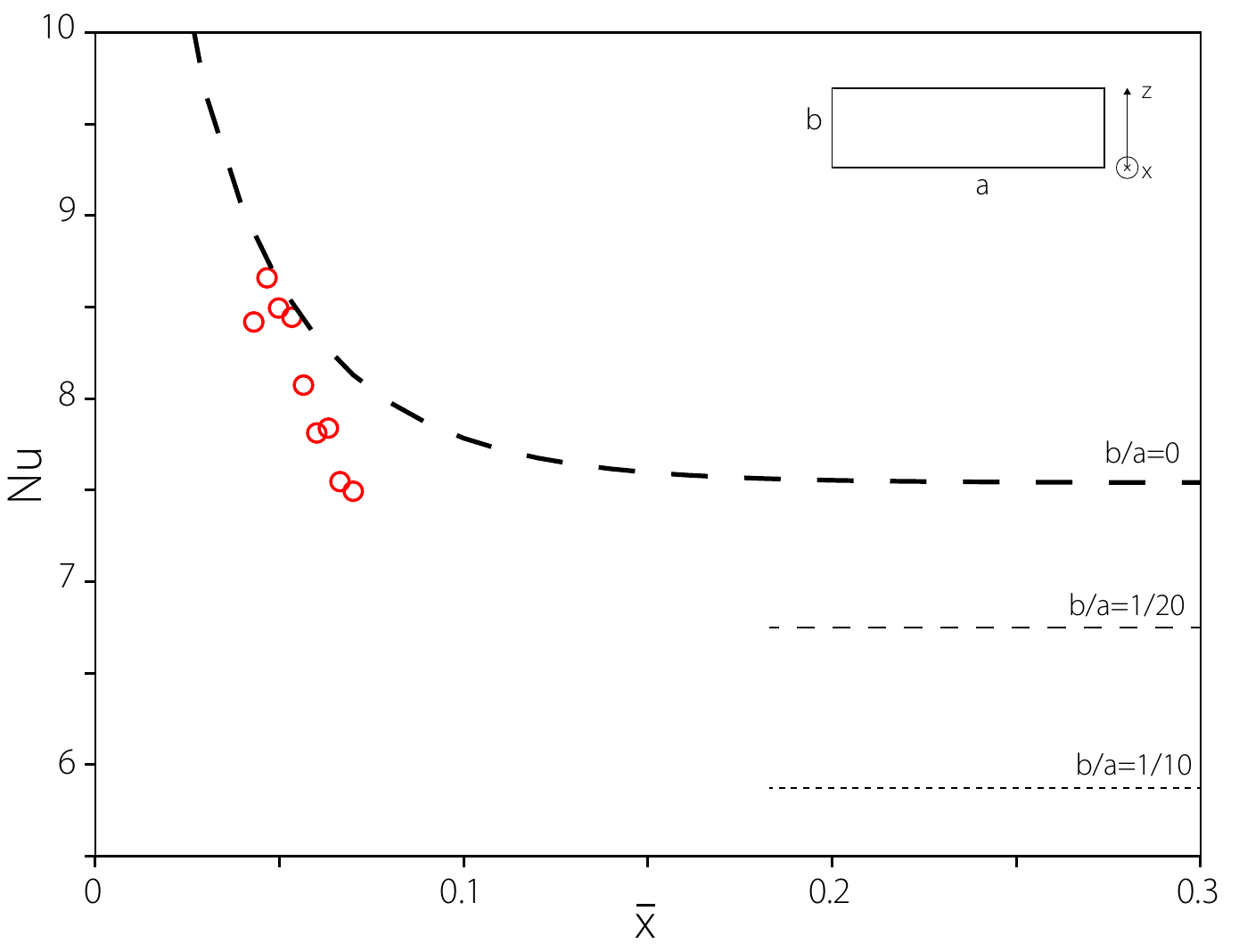}
  \caption{Nusselt number as a function of the normalised positions on the plane. \textcolor{red}{$\circ$}~:~Experimental data points. 
   \protect\rule[0.5ex]{0.25cm}{0.4mm}\hspace{0.25cm}\protect\rule[0.5ex]{0.25cm}{0.4mm} : Theoretical curve derived from the theoretical water temperature field;
   \protect\rule[0.5ex]{0.1cm}{0.2mm}\hspace{0.1cm}\protect\rule[0.5ex]{0.1cm}{0.2mm}\hspace{0.1cm}\protect\rule[0.5ex]{0.1cm}{0.2mm} : Asymptotic value of $Nu$ found for a rectangular duct of aspect ratio 1/20;
     \protect\rule[0.5ex]{0.5mm}{0.2mm}\hspace{0.5mm}\protect\rule[0.5ex]{0.5mm}{0.2mm}\hspace{0.5mm}\protect\rule[0.5ex]{0.5mm}{0.2mm}\hspace{0.5mm}\protect\rule[0.5ex]{0.5mm}{0.2mm}\hspace{0.5mm}\protect\rule[0.5ex]{0.5mm}{0.2mm} : Asymptotic value of $Nu$ found for a rectangular duct of aspect ratio 1/10.
  }
\label{fig:Nusselt}
\end{figure}

Figure~\ref{fig:Nusselt} presents with a dashed line (thick) the theoretical prediction of the Nusselt number $Nu$ derived from the model presented above. As expected from the symmetry of our model, the $Nu$ dashed line curve converges toward a value of 7.54, which compares well to the value obtained for a fully-developed laminar flow in a Hele-Shaw cell \citep{Incropera2007}. 
The red empty dots show the corresponding experimental estimations of $Nu$. We recall that in this estimation the velocity profile appearing in $<T>$ is not measured, and that we used the half-Poiseuille field of the theoretical analysis. The experimental values of the Nusselt number do not correspond perfectly to this two dimensional prediction, although the tendency and the order of magnitude are correct. 
In fact, we can adapt our 2D model to obtain predictions for a three dimensional rectangular duct, where the rivulet width would be denoted $a$ and its thickness ($h_{\rm w}$ for us) denoted $b$ as shown in the schematic diagram on the top right corner of Figure~\ref{fig:Nusselt}. Our 2D model corresponds thus to the null aspect ratio ($b/a=0$).
The horizontal dashed lines (thin) represent the known asymptotic values found for such ducts with an aspect ratio of $b/a=20$ and $10$, from top to bottom respectively \citep{Shah2014}.
Even though we do not clearly reach any asymptotic regime for $Nu$ in our experiment, we can forecast a plateau value for $Nu$ between 6 and 7. This value is close to the case of a duct with an aspect ratio between 10 and 20. It shows that in our experiment, we have a small effect of the three dimensional geometry of the rivulet, reducing slightly the efficiency of the heat transfer as compared to the ideal 2D model.

\subsubsection{Surface temperature}

\begin{figure}[h!]
\centering
  \includegraphics[width=1\textwidth]{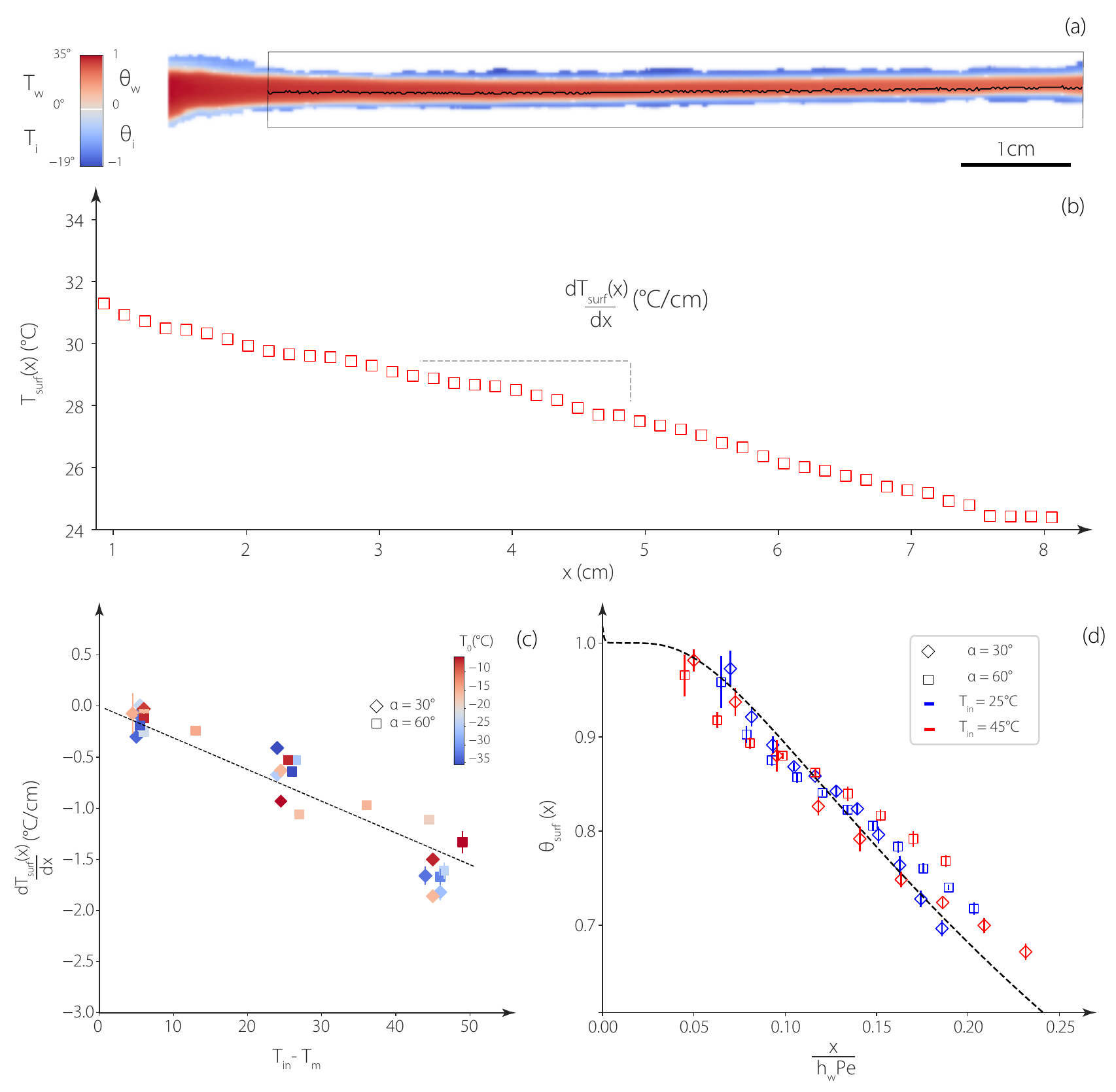}
  \caption{Surface temperature of the flowing water. (a) Thermal picture of an  experiment with $T_{\rm in} = 35^{\circ}$C, and $T_{\rm s} = -19^{\circ}$C. The colour-bar corresponds to the temperatures in the ice and in the water. (b) Surface temperature measured inside the black square following the black line at the centre. (c) Temperature gradients of all the conducted experiments against $T_{\rm in}$. Markers stand for the angle inclination and the ice-substrate interface temperature $T_{\rm 0}$ is colour-coded. The dashed line represents the best fit of the data. (d) Rescaled surface temperature fields $\theta_{\rm surf}$ for 2 different injection temperatures and inclinations as a function of the rescaled plane position. $T_{\rm in}$ is taken as the maximum value measured by the camera close to the needle. The dashed line is the theoretical prediction $\theta_{\rm surf}(\bar{x})=\theta_w(\bar{x},1)$.}
\label{fig:Tsurf}
\end{figure}

Figure~\ref{fig:Tsurf}(a) shows a typical experimental temperature field measured from above. The water is injected on the left and flows downward, to the right. The image is taken in the permanent regime. As indicated on the colour bar on the left of this temperature field, the positive temperatures (water) are colour-coded in red whereas the negatives ones (ice) are colour-coded in blue.
It is first interesting to notice that we observe a layer of ice wider than the water layer. A possible explanation for this effect is the difference of wetting properties of water on metal and on ice \citep{Thievenaz2020}. Before freezing, the water rivulet has a given width on the metal \citep{Towell1966}, then it freezes and an ice structure grows on the metal. The initial ice layer appears in blue on the temperature field and its width corresponds to the width of the initial rivulet.
During the solidification process, the water retracts while flowing to reach a constant width, as shown in red on the picture. The width of a rivulet being a balance between the flow-rate, the surface tension and the contact angle, it suggests that there is an evolution of the contact angle during the experiment \citep{Thievenaz2020}.
We also notice in the picture a small entrance zone where the width varies along the flow. This is a direct signature of the injection process with the needle. In the following, the measurements are taken out of this area, in the zone between 2 and 9\,cm, materialised by the rectangle in the picture. Remarkably, the length of this transient region is similar to the characteristics growth scale of the viscous and thermal boundary layers, so that we expect our theoretical analysis to be pertinent only outside of this domain.

Thus, equation (\ref{ap:thetaw}) can be used at the surface ($\bar{z}=1$, that is $z=h_{\rm w}$) to obtain an approximation for the surface temperature far from the needle:
\begin{equation}
    T_{\rm surf}(x)=T_{\rm m}+(T_{\rm in}-T_{\rm m})\,A_1\,\Phi_1(1)\,\exp\left(-\lambda_1^2\frac{x}{h_{\rm w}\,Pe}\right)
    \label{eq:Tsurf}
\end{equation}
After linearisation, we obtain a linear variation of the surface temperature along the plane ($x$-axis) and we deduce a theoretical expression for its slope:
\begin{equation}
    \frac{{\rm d}T_{\rm surf}}{{\rm d}x}=-(T_{\rm in}-T_{\rm m})A_1\,\Phi_1(1)\,\frac{\lambda_1^2}{h_{\rm w}Pe}
    \label{eq:dTsurf}
\end{equation}
Interestingly, the model predicts that the slope is linear with the injection temperature of water $T_{\rm in}$ and does not depend on the substrate temperature $T_{\rm 0}$.

In order to compare this prediction to our measurements, we extract experimentally the surface temperature. We define it as the maximum value for each $x$-position on the plane, and it gives the black line drawn in the picture. The corresponding profile is plotted in Figure~\ref{fig:Tsurf}(b) and shows the linear decrease of the surface temperature along the plane predicted by the model. This linear decrease is retrieved for all experiments.
Figure~\ref{fig:Tsurf}(c) presents the slope of these surface temperature profiles as a function of the inlet water temperature $T_{\rm in}-T_{\rm m}$ for various substrate temperatures $T_0$ and two different angles $\alpha$. We observe that the slope is naturally stronger for higher injection temperature of water and we recover the linear trend with $T_{\rm in}$ predicted by equation (\ref{eq:dTsurf}). The dashed line is a linear fit of the data with a prefactor $-3.2 \,$m$^{-1}$ that is compatible with the prediction ($-7.4 \, $m$^{-1}$). We attribute this discrepancy mainly to the arbitrary choice of the velocity $U_0$ that is present in the definition of the Peclet number and which has been determined within our 2D model.
Furthermore, we also show with this plot that the decrease of the surface temperature of water is independent of the temperature of the substrate-ice interface $T_0$, as it was predicted by equation (\ref{eq:dTsurf}). This emphasises the view exposed in the theoretical part: the temperature fields in both the ice and the water are disconnected, leaving the ice thickness as the only parameter ensuring the heat flux continuity. 

Finally, Figure~\ref{fig:Tsurf}(d) shows four normalised experimental temperature surface profiles $\theta_{\rm surf}$ obtained for two different angles $\alpha$ and two different injection temperatures $T_{\rm in}$, plotted as a function of $x$ normalised by $h_{\rm w}\,Pe$. We recover the linear variation observed in (b). 
On the same plot, the black dashed line is the full prediction of the model for the rescaled surface temperature $\theta_w(\bar{x},1)$ as already plotted as a solid line in Figure~\ref{fig:Tmap}. Once again, there is a very good agreement between the theoretical prediction and the experimental results, confirming the suitability of the model to characterise the heat exchange in the frozen rivulet.

\section{Conclusion}

In this paper we have analysed experimentally and theoretically the formation, growth, final steady shape and temperature fields of a freezing rivulet resulting from a water thread flowing down a cold solid plate. We performed experiments varying the inclination of the plate, its temperature, and the water injection temperature.

The model developed is based on the resolution of the heat equations in the ice (diffusion) and in the water (advection-diffusion).
In the ice, we found a temperature field varying linearly from the substrate-ice interface temperature to the ice-water interface one.
In the water, however, the solution is the superposition of different cylindric functions which is well approximated by its first term to describe the temperature field far from the injection needle.

Interestingly, due to the injection condition (constant temperature), a thermal boundary layer develops, responsible for a temperature gradient in the water and thus a thermal flux at the ice-water interface. This thermal boundary layer thickens with the distance from the needle and reaches the free surface of the rivulet, few centimetres downstream. Once the boundary layer has established the heat flux in the water is in fact constant over time. 
The Stefan boundary condition then governs the ice layer growth rate as long as the thermal fluxes from both sides of the ice-water interface are not equal.
This is what we observed experimentally with the existence of three different regimes over time. 

In the first one, the ice layer growth is homogeneous along the plane and evolves with the square-root of time. This regime is described in a previous paper \citep{Monier2019} and well explained by considering the Stefan boundary condition neglecting the thermal flux in the water.
However, increasing the ice layer thickness reduces the thermal flux in the ice and leads inevitably to a second regime where both heat fluxes have to be considered.

We found experimentally that, in this regime, the ice layer thickness converges towards a maximum with an exponential relaxation rate over time. The water temperature is found to be the dominant parameter to control this convergence time, in good agreement with the model prediction.
Finally, when the heat flux in the ice becomes equal to the one in the water, a permanent regime is reached. 

There, the ice layer adopts a striking structure, showing a linear thickening with the plane position, located after a small transient region close to the needle where the ice layer is more abrupt. This ice shape is well captured by our model.
In order to fully characterise this permanent regime, we used an infrared camera to record the vertical temperature fields in the ice and in the water. 
The good agreement between theses measurements and the developed model highlights the link between the temperature field in the water and the ice layer shape. 
In a first zone, close to the injection needle, where the thermal boundary layer thickens, the ice layer profile shows a steep increase. After few centimetres, the thermal boundary layer reaches the water-air interface and the water surface temperature starts to decrease along the plane. 
Measurements of the surface temperature confirm a linear temperature decrease and the model highlights the clear link between this linear decrease and the linear shape of the ice layer. This link is further confirmed by the very good agreement between the model predictions for the geometrical features and the experimental data. 

It is important to emphasise that in our approach no adjustable parameter nor macroscopic heat exchange coefficient were needed to account theoretically for the experimental results. 
Overall, by quantitatively comparing the experiments and the model, our work demonstrates that the dynamics and the final state of a freezing rivulet can be totally determined by the precise balance between the quasi-static thermal fluxes in both domains, water and ice.

\section*{Acknowledgements} 

The authors would like to warmly acknowledge P.-Y. Lagr\'ee for fruitful discussions and the Direction G\'en\'erale de l'Armement (DGA) for financial support.

\section*{Appendix}

The general solution $\Phi_n$ can be written as a sum of two functions $\varphi_{1n}$ and $\varphi_{2n}$:

\begin{equation}
\Phi_n(z,\lambda_n)= c_n\left(\varphi_{1n}(z,\lambda_n)-\frac{\varphi_{1n}(0,\lambda_n)}{\varphi_{2n}(0,\lambda_n)}\varphi_{2n}(z,\lambda_n)\right)
\end{equation}
\\
where $c_n$ is a real constant and the functions $\varphi_{in}$ are defined as:

\begin{equation}
\varphi_{1n}(z,\lambda_n) =   \Re\left(D_{\frac{-\lambda_n-1}{2}}\left(i \sqrt{2\lambda_n}( z-1)\right)\right)  \; \textrm{and}\; \varphi_{2n}(z,\lambda_n)= \Re\left(D_{\frac{\lambda_n-1}{2}}\left(\sqrt{2\lambda_n}(z-1)\right)\right)
\end{equation}
\\
where the functions $D_k(z)$ are parabolic cylinder functions \citep{whittaker1996course}. Table \ref{tab:lbdn} presents the nine first non-zero roots $\lambda_n$ with the associated coefficient $c_n$ and $A_n$.
 \begin{table}[h!]
  \centering
 \begin{tabular}{|cccc|}
              	  n & $\lambda_n$ & $c_n$ & $A_n$\\[3pt]\hline
                  1 & 1.6816 & 1.28245 & 0.779039 \\[3pt]
                  2 & 5.66986 & 3.0843 & -0.188224 \\[3pt]
                  3 & 9.66824 & 13.3959 & 0.100925 \\[3pt]
                  4 & 13.6677 & 84.8862 & -0.0673737 \\[3pt]
                  5 & 17.6674 & 707.515 & 0.0499319 \\[3pt]
                  6 & 21.6672 & 7311.54 & -0.0393499 \\[3pt]
                  7 & 25.6671 & 90178.6 & 0.0322913 \\[3pt]
                  8 & 29.667 & 1.29258$\cdot$10$^6$ & -0.0272704 \\[3pt]
                  9 & 33.667 & 2.11122$\cdot$10$^7$ & 0.0235286 \\[3pt]\hline
      \end{tabular}
      \caption{Values of $\lambda_n$, $c_n$, and $A_n$ for the non-trivial solutions.}   \label{tab:lbdn}          
\end{table}

\bibliographystyle{jfm}
\bibliography{rivulet}

\end{document}